\documentclass[%
 reprint,
 showpacs,
 amsmath,amssymb,
 aps,
 pra,
]{revtex4-1}

\usepackage[ascii]{inputenc}
\usepackage[T1]{fontenc}
\usepackage[english]{babel}
\usepackage{graphicx}
\usepackage{dcolumn}
\usepackage{bm}
\usepackage{mathtools}

\usepackage{hyperref}

\DeclareMathOperator{\imag}{Im}
\DeclareMathOperator{\real}{Re}
\newcommand{\PT}{\mathcal{PT}}
\newcommand{\mbf}{\mathbf}
\newcommand{\mrm}{\mathrm}
\newcommand{\rmi}{\mathrm{i}}
\newcommand{\rme}{\mathrm{e}}
\newcommand{\rmd}{\mathrm{d}}
\newcommand{\rmk}{\mathrm{k}}

\newcommand{\ket}[1]{\left|{#1}\right\rangle}
\newcommand{\braket}[2]{\left\langle{#1}\middle|{#2}\right\rangle}

\begin{document}

\preprint{PRA/PT BEC}

\title{Nonlinear quantum dynamics in a $\PT$-symmetric double well}

\author{Daniel Haag}
\email[]{Daniel.Haag@itp1.uni-stuttgart.de}

\author{Dennis Dast}

\author{Andreas L\"ohle}

\author{Holger Cartarius}

\author{J\"org Main}

\author{G\"unter Wunner}

\affiliation{Institut f\"ur Theoretische Physik 1,
  Universit\"at Stuttgart, 70550 Stuttgart, Germany}

\date{\today}

\begin{abstract}
  We investigate the mean-field dynamics of a Bose-Einstein condensate (BEC)
  described by the Gross-Pitaevskii equation (GPE) in a double-well potential
  with particle gain and loss, rendering the system $\PT$-symmetric.
  The stationary solutions of the system show a change from elliptically stable
  behavior to hyperbolically unstable behavior caused by the appearance of
  $\PT$-broken solutions of the GPE and influenced by the nonlinear
  interaction. The dynamical behavior is visualized using the Bloch sphere
  formalism. However, the dynamics is not restricted to the surface of the
  sphere due to the nonlinear and non-Hermitian nature of the system.
\end{abstract}


\maketitle


\section{Introduction}
The interaction with the environment often plays an important role in
studies of ultracold atoms, leading to gain or loss of particles. One important
example is the inelastic three-body collision of particles in a BEC which can
be described in mean-field approximation by an imaginary interaction potential,
thus rendering the Hamiltonian non-Hermitian~\cite{Moiseyev2011a, Kagan98a}.
Imaginary potentials also find application in studies of dissipative
optical lattices~\cite{Abdullaev10a, Bludov10a}, and the non-Hermitian GPE has
been derived as the mean-field limit of an open Bose-Hubbard system described
by a master equation in Lindblad form~\cite{Trimborn08a, Witthaut11a}. A gain
of particles is less described in literature, but the feeding of a condensate
from a thermal cloud has been described by a positive imaginary
potential~\cite{Kagan98a}.

We study the dynamics of a BEC in a double-well potential where particles are
coherently removed from one well and injected into the other. The system
is described in mean-field approximation by the GPE. The GPE is known to yield
accurate results for temperatures considerably smaller than the critical
temperature of the condensate but has limitations in the vicinity of dynamic
instabilities~\cite{Anglin01a, Vardi01a}. The removal and injection of
particles is described by an imaginary potential. Both a coherent influx and
outflux have been experimentally realized. A coherent particle loss has, e.g.,
been implemented using a focused electron beam~\cite{Gericke08a} whereas the
influx can be provided from a second condensate exploiting electronic
excitations of the atoms~\cite{Robins08a}.

The gain and loss contributions are chosen in such a way that the resulting
system is $\PT$-symmetric, where $\mathcal{P}$ denotes the parity operator
$\hat{x}\to-\hat{x}$, $\hat{p}\to-\hat{p}$ and $\mathcal{T}$ the time reversal
operator $\hat{p}\to-\hat{p}$, $\rmi\to-\rmi$. Even though being non-Hermitian,
such systems can exhibit entirely real eigenvalues in certain parameter
regimes~\cite{Bender98a}. In recent years efforts have been pursued to
establish a quantum theory in which the requirement of Hermiticity is replaced
by the weaker condition of $\PT$ symmetry~\cite{Bender99a, Bender07a} or by the
more general concept of pseudo-Hermiticity~\cite{Mostafazadeh02a,
Mostafazadeh02b, Mostafazadeh02c}. Besides these fundamental approaches to
generalize quantum mechanics a great variety of $\PT$-symmetric systems which
are potentially experimentally accessible has been theoretically
investigated~\cite{Klaiman08a, Schindler11a, Bittner12a, Cartarius12a,
Mayteevarunyoo13a, Graefe12b}. The experimental breakthrough succeeded in
optical waveguide systems~\cite{Klaiman08a, Rueter10a, Guo09a}, now being one
of the main foci in the field of $\PT$ symmetry.

For the system considered $\PT$ symmetry demands that the influx and outflux of
particles into or from the condensate is balanced in such a way that
stationary solutions can be found. Embedding the $\PT$-symmetric double well
into a Hermitian four-well potential is a possible experimental realization of
this system~\cite{Kreibich13a}.

Prior to the experimental realization of a $\PT$-symmetric nonlinear quantum
system it is of utmost importance to have a detailed understanding of the
dynamical behavior resulting from the combination of $\PT$ symmetry and
nonlinearity. In particular it is necessary to study the implications on the
stability properties of the stationary solutions since only stable states are
observable. However, not only the stability of stationary states but also the
dynamics of arbitrary wave packets is relevant because for certain initial wave
packets the number of particles diverges, thus destroying the condensate. We
will address these problems, laying the foundation for future attempts of an
experimental realization of a $\PT$-symmetric Bose-Einstein condensate.

In this article we solve the dimensionless GPE with contact interaction
\begin{equation}
  \left[-\Delta+V(\mbf{r})+8\pi N_0 a|\psi(\mbf{r},t)|^2\right] \psi(\mbf{r},t)
  =\rmi \frac{\partial}{\partial t} \psi(\mbf{r},t),
  \label{eq:gpe}
\end{equation}
where $N_0$ is the number of particles and $a$ is the scattering length. The
three-dimensional double-well potential chosen,
\begin{equation}
  V(\mbf{r})= \frac14 x^2 + \frac14 \omega_{y,z}^2(y^2+z^2)
  + v_0 \rme^{-\sigma x^2} + \rmi \gamma x \rme^{-\rho x^2},
  \label{eq:potential_3d}
\end{equation}
is $\PT$-symmetric since $V(\mbf{r})=V^*(-\mbf{r})$ holds. The
potential consists of a three-dimensional harmonic trap with an identical
trapping frequency $\omega_{y,z}$ in $y$ and $z$ direction. The harmonic trap
in $x$ direction is superimposed by a Gaussian barrier with height $v_0$ and
width parameter $\sigma$, thus forming a symmetric double-well potential. The
strength of the antisymmetric imaginary part of the potential is tuned by
$\gamma$, and the parameter $\rho$ is chosen in such a way that the extrema of
the imaginary part coincide with the minima of the double well. The positive
imaginary part can be interpreted as a source of probability density while a
negative part corresponds to a sink. In all following calculations the
values $\omega_{y,z}=2$, $v_0=4$, $\sigma=0.5$, and $\rho\approx0.12$ are fixed
whereas the gain/loss parameter $\gamma$ and the strength of the nonlinearity
$N_0a$ are varied.

Since the system is non-Hermitian the norm of the wave function $\psi$ is not
conserved and due to the nonlinearity of the GPE a different norm changes the
dynamics. The physical interpretation of a change in the norm of the wave
function is a change in the number of particles
\begin{equation}
  N = N_0 ||\psi||^2.
  \label{eq:particlenumber}
\end{equation}

Without interaction, i.e.\ $Na=0$, the Hamiltonian is separable and is solved
by the product ansatz $\psi(x)\psi_m(y)\psi_m(z)$, where $\psi_m$ is the $m$-th
eigenstate of the one-dimensional harmonic oscillator. Since the energy of the
first excited state of the harmonic oscillator is about one order of magnitude
larger than that of the first excited state of the double well, only the ground
states of the harmonic oscillators in $y$ and $z$ direction are taken into
account. Thus the three-dimensional problem is reduced to one dimension with
the one-dimensional $\PT$-symmetric double-well potential
\begin{equation}
  V(x)= \frac14 x^2 + v_0 \rme^{-\sigma x^2} + \rmi \gamma x \rme^{-\rho x^2}.
  \label{eq:potential_1d}
\end{equation}
This can easily be solved by a numerically exact integration.
The product ansatz does not exactly solve the nonlinear GPE with
contact interaction, however, for small values of $Na$ it is still a good
approximation, and as we will see not only shows qualitatively the same
behavior as the calculations in three dimensions but also quantitatively.

The article is organized as follows. In Sec.~\ref{sec:stationary} the
stationary solutions in three dimensions are presented and a comparison with
the one-dimensional solutions is drawn.
To study the stability of the stationary solutions the Bogoliubov-de Gennes
equations are adapted for non-Hermitian systems and solved numerically in
Sec.~\ref{sec:stability}. In Sec.~\ref{sec:dynamics} the time evolution
of wave packets is investigated using the Bloch sphere formalism. Conclusions
are drawn in Sec.~\ref{sec:conclusion}.

\section{Stationary solutions}
\label{sec:stationary}
The eigenvalue spectrum and eigenstates of a BEC in the $\PT$-symmetric double
well~\eqref{eq:potential_1d} have already been discussed~\cite{Dast12a,
Dast13a}, however, only as a function of the gain/loss parameter $\gamma$. To
study dynamical properties it is more instructive to discuss the eigenvalues as
a function of $Na$, where $N$ is the effective number of particles as defined
in Eq.~\eqref{eq:particlenumber}. 

To solve the three-dimensional system we use the \textit{time-dependent
variational principle}~\cite{Mclachlan64a,Rau2010c,Rau10a,Rau10b} whose
application to $\PT$-symmetric nonlinear systems has been discussed
in~\cite{Dast12a}. Our ansatz consists of coupled Gaussian functions
\begin{align}
  \psi(\mbf{r}) = \sum_{\rmk}
  \exp \big[ -A_x^\rmk (x-q_x^\rmk)^2 + A_{y,z}^\rmk (y^2+z^2) \notag\\
  - \rmi p_x^\rmk (x-q_x^\rmk) + \varphi^\rmk \big]
  \label{eq:tdvp_ansatz}
\end{align}
with $A_x^\rmk, A_{y,z}^\rmk, \varphi^\rmk \in \mathbb{C}$, $q_x^\rmk, p_x^\rmk
\in \mathbb{R}$. 
The variational principle yields equations of motion for these time-dependent
quantities, and stationary solutions are found as the fixed points of these
equations. The stationary solutions of the three-dimensional
potential~\eqref{eq:potential_3d} presented in this section are gained with
the variational approach whereas the one-dimensional
potential~\eqref{eq:potential_1d} is solved numerically exact. We use up to
two Gaussian functions per well, i.e.\ up to four Gaussians in total, which
yields a small correction to one Gaussian function per well.

\subsection{Spectrum of the real double well}
\begin{figure}%
  \centering%
  \includegraphics[width=\columnwidth]{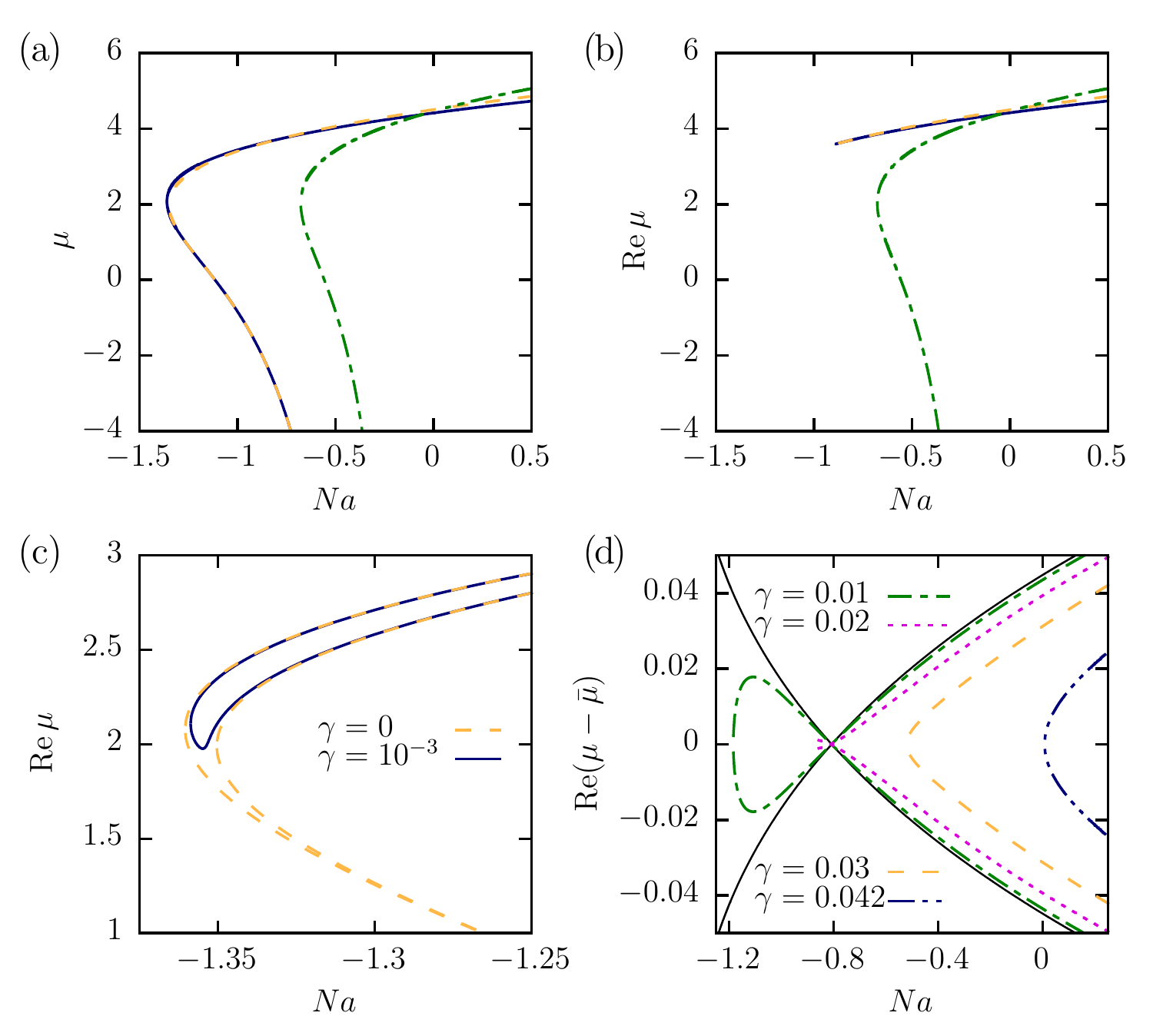}%
  \caption{%
    (Color online)
    (a) The chemical potential $\mu$ of the eigenvalues without gain/loss,
    i.e.\ $\gamma=0$, as a function of the particle number scaled scattering
    length. The symmetric (blue solid branch) and antisymmetric (yellow dashed
    branch) stationary solutions arise at two almost identical bifurcation
    points at $Na\approx-1.4$, and the doubly degenerate symmetry-breaking
    eigenstates (green dot-dashed branch) at $Na\approx-0.7$. (b) Introducing
    the gain/loss $\gamma=0.02$ hardly changes $\real \mu$ of the
    symmetry-breaking ($\PT$-broken) eigenstates but the two $\PT$-symmetric
    branches now arise at a tangent bifurcation. (c) The tangent bifurcation of
    the two $\PT$-symmetric states in the vicinity of the bifurcation point for
    a small value of $\gamma$. (d) The difference between the chemical
    potential of the $\PT$-symmetric states and their mean value $\bar{\mu}$.
    Different values of $\gamma$ are compared to the case $\gamma=0$ (black
    solid line).
  }%
  \label{fig:spectrum_3d}%
\end{figure}

Figure~\ref{fig:spectrum_3d} shows the eigenvalues of the GPE in the
three-dimensional $\PT$-symmetric double well~\eqref{eq:potential_3d} with the
lowest total energy. For $\gamma=0$ the spectrum in
Fig.~\ref{fig:spectrum_3d}(a) contains three pairs of states each of which form
a common branch, shown as blue solid, yellow dashed and green dot-dashed lines.
Note that the blue and yellow branches almost lie on top of each other. Every
branch is born in a tangent bifurcation at a critical value of $Na$. At the
three tangent bifurcations not only the eigenvalues but also the eigenstates
coincide. 

The two bifurcations for the blue and yellow branches reside at almost
identical values $Na \approx -1.4$. The states arising at these tangent
bifurcations have an equal probability of presence in both wells. At one
bifurcation two states with symmetric (even parity) wave functions arise (blue
solid lines), whereas the two states arising at the other bifurcation (yellow
dashed lines) are antisymmetric (odd parity). At the bifurcation points the
wave functions turn into two peaks strongly confined in the two wells due to
the attractive interaction. For stronger attractive interactions no stationary
solutions exist and the condensate collapses~\cite{Gammal01a, Donley01a}.

\begin{figure}%
  \centering%
  \includegraphics[width=\columnwidth]{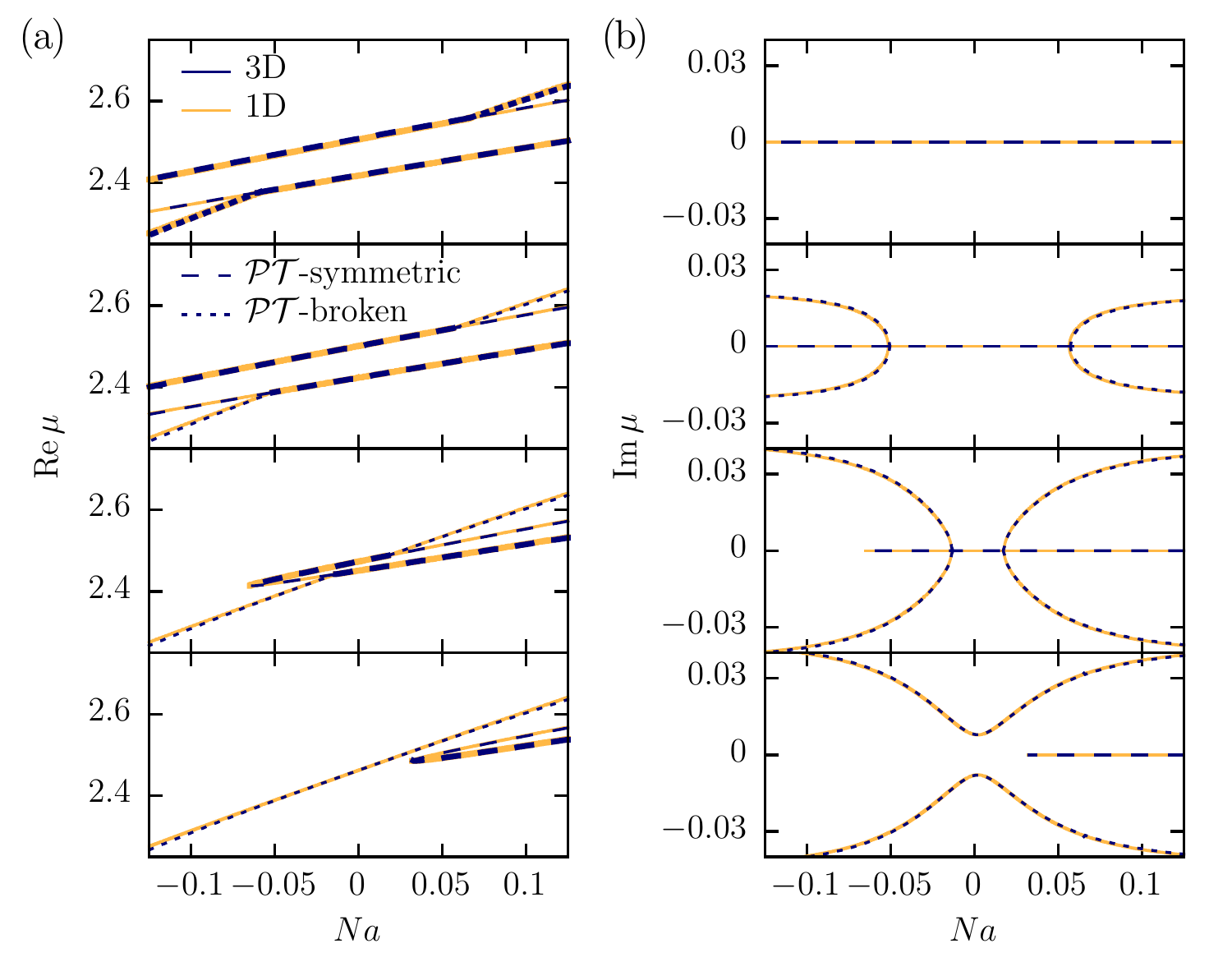}%
  \caption{%
    (Color online)
    (a) Real and (b) imaginary part of the chemical potential of the
    eigenstates for small values of $Na$. From top to bottom the four different
    gain/loss contributions $\gamma=0$, $0.02$, $0.04$, and $0.042$ are used.
    The eigenvalues of the $\PT$-symmetric and $\PT$-broken solutions in three
    dimensions are in excellent agreement with the numerically exact
    one-dimensional solutions. Stable branches are highlighted as thick
    lines (see Sec.~\ref{sec:stability}).
  }%
  \label{fig:spectrum}%
\end{figure}

The same collapse is observed at approximately half the interaction strength
$Na \approx -0.7$ (green dot-dashed branch) where the wave function is entirely
localized in one well, i.e.\ it corresponds to exactly the same collapse
process of the wave function, but for an asymmetric wave restricted to one side
of the double well.  The states arising at this bifurcation are clearly parity
symmetry broken and doubly degenerate because the condensate can be located
either in the right or in the left well. The two uppermost panels of
Fig.~\ref{fig:spectrum} shows the eigenvalues for small moduli of $Na$ in
more detail revealing that the parity symmetry broken states coalesce with the
symmetric solution at $Na\approx-0.0075$ and with the antisymmetric solution at
$Na\approx0.0075$. In the region $|Na|\lesssim0.0075$ the parity symmetry
breaking solutions do not exist. The existence of these solutions is known as
macroscopic quantum self-trapping~\cite{Albiez05a}.

\subsection{Spectrum of the $\PT$-symmetric double well}
\begin{figure}%
  \centering%
  \includegraphics[width=\columnwidth]{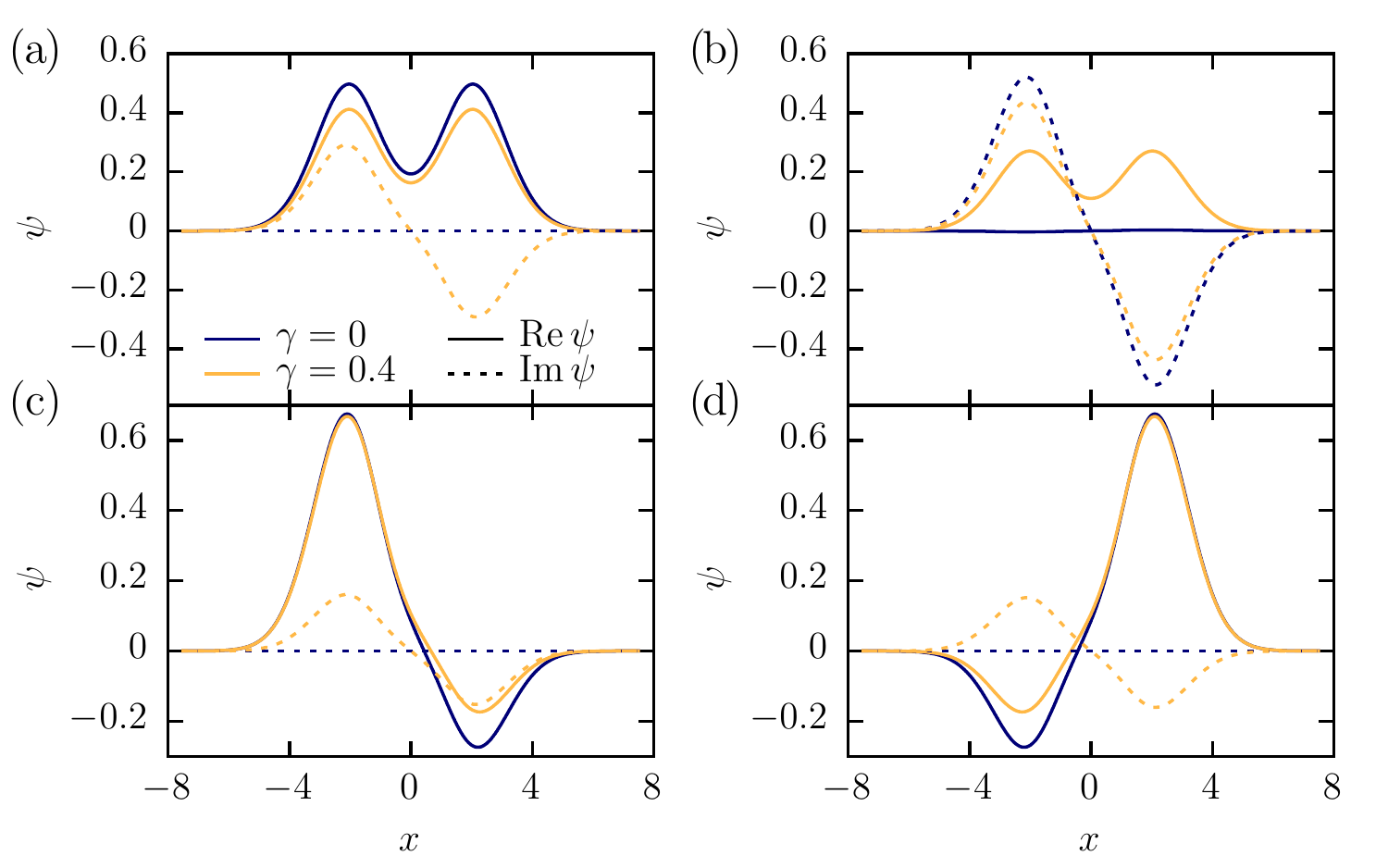}%
  \caption{%
    (Color online)
    Real (solid lines) and imaginary (dotted lines) part of the wave function
    of the $\PT$-symmetric ground (a) and first excited (b) state and the two
    $\PT$-broken states with a negative (c) and positive (d) imaginary part of
    the chemical potential. For $\gamma=0$ the ground state is parity symmetric
    and the first excited state is antisymmetric. For finite values of $\gamma$
    the real part of both $\PT$-symmetric wave functions is even and the
    imaginary part is odd. The $\PT$-broken states are asymmetric thus breaking
    the $\PT$ symmetry of the Hamiltonian.
  }%
  \label{fig:wavefnc}%
\end{figure}

Introducing the gain and loss contribution $\gamma=0.02$ changes the eigenvalue
spectrum significantly as can be seen in Fig.~\ref{fig:spectrum_3d}(b). The
even and odd state with lower energies vanish and the two
remaining states only exist for $Na\gtrsim-0.9$. These two states are
$\PT$-symmetric, thus having a symmetric real part and an antisymmetric
imaginary part (see Fig.~\ref{fig:wavefnc}(a),(b)). For the symmetry-breaking
states the real part of the chemical potential is approximately identical to
the case $\gamma=0$ but the eigenvalue $\mu$ is now complex. Due to the $\PT$
symmetry of the system, the two imaginary parts of $\mu$ have the same absolute
value but different signs. The wave functions of the $\PT$-broken states are
shown in Fig.~\ref{fig:wavefnc}(c),(d). To understand the behavior of the two
tangent bifurcations of the symmetric and antisymmetric states we compare the
vicinity of the bifurcation points for $\gamma=0$ and $\gamma=10^{-3}$.
Figure~\ref{fig:spectrum_3d}(c) shows that the two bifurcations vanish and a
new tangent bifurcation between the two $\PT$-symmetric states arises.

If we want to compare the eigenvalues of the two $\PT$-symmetric states for
different values of $\gamma$ we have to choose a slightly different
presentation since the difference between the eigenvalues of the two
$\PT$-symmetric states is very small compared to their absolute change.
Figure~\ref{fig:spectrum_3d}(d) shows $\mu-\bar{\mu}$ of the two
$\PT$-symmetric states, where $\bar{\mu}$ denotes the mean value of $\mu$ of
these two states. We immediately see that the tangent bifurcation at which the
two $\PT$-symmetric states coalesce is shifted to greater values $Na$ if the
gain/loss parameter $\gamma$ is increased. For $\gamma\gtrsim0.41$ the
bifurcation point is shifted to positive values of $Na$, i.e.\ the two
$\PT$-symmetric solutions investigated here only exist for repulsive
interaction of the atoms. At $Na\approx-0.8$ the chemical potential of the two
$\PT$-symmetric states becomes equal for all parameters $\gamma$ for which the
$\PT$-symmetric states exist at $Na\approx-0.8$. However, at this point only a
generic degeneracy of the energy eigenvalue $\mu$ occurs, the states
themselves are different.

The relevant $\PT$-symmetric effects of the system, namely the breaking of
$\PT$-symmetry, occur already at small absolute values of $Na$. This regime
is shown in Fig.~\ref{fig:spectrum}. The $\PT$-broken states emerge from the
$\PT$-symmetric state in a pitchfork bifurcation. For an attractive
interaction the $\PT$-broken solutions emerge from the $\PT$-symmetric ground
state whereas in the case of repulsive interaction the $\PT$-broken branches
emerge from the excited $\PT$-symmetric state (cf.\ middle two panels in
Fig.~\ref{fig:spectrum}). The $\PT$-broken solutions arise at approximately the
same absolute value of $Na$ for both attractive and repulsive interaction. As
is known from $\PT$-symmetric systems the eigenvalues of the $\PT$-broken
solutions occur in complex conjugate pairs. For increasing values of the
gain/loss parameter $\gamma$ the tangent bifurcation in which the two
$\PT$-symmetric states vanish is shifted to greater values of $Na$. At the same
time the $\PT$-broken solutions emerge already at smaller absolute values of
$Na$. At strong enough values of $\gamma$ the tangent bifurcation is shifted
to repulsive interactions $Na>0$ and the $\PT$-broken states exist even for
$Na=0$.

Although being solutions of the time-independent GPE the $\PT$-broken
states are no stationary solutions of the time-dependent GPE. They
experience an exponential gain or decay of the norm, thus effectively changing
the nonlinearity parameter $Na$.

In the following we will restrict the discussion to small moduli of $Na$. As
already mentioned in this regime the GPE is in good approximation solved by a
product ansatz, thus reducing the problem to one dimension. The comparison in
Fig.~\ref{fig:spectrum} confirms the excellent agreement between the
calculations in one and three dimensions for the parameter range considered and
justifies the reduction to one dimension used in the following sections.
The solutions for the one-dimensional
system~\eqref{eq:potential_1d} are obtained with numerically exact methods by
integrating the GPE outwards and fulfilling boundary conditions \cite{Dast12a}.

\section{Stability}
\label{sec:stability}
The first step towards understanding the dynamical properties of the system is
the stability analysis of the stationary solutions with respect to small
perturbations. The stability of the $\PT$-symmetric stationary states has
already been discussed rudimentarily in~\cite{Dast12a}. We will shortly review
these results and then focus on the stability in the vicinity of the
bifurcations and the study of the dynamics of $\PT$-broken solutions.

In addition to the stability analysis in one dimension, which is presented in
this section, we investigated the stability in three dimensions by linearizing
the equations of motion of the time-dependent variational principle as
described in~\cite{Dast12a}. In three dimensions excitations in $y$ and $z$
direction may give rise to additional instabilities. However, we found that in
the parameter range considered these instabilities do not occur for
$\omega_{y,z} \gtrsim 3$ and we again observe an excellent agreement between
the calculations in one and three dimensions.

The time-dependent GPE is linearized in the vicinity of the stationary states,
yielding the Bogoliubov-de Gennes equations
\begin{subequations}
\begin{align}
  \Delta u=&
   \left(V-\mu-\omega-8N_0a\left|\psi_0\right|^2\right)u-4N_0a\psi_0^2v,\\
  \Delta v=&
   \left(V^*-\mu^*+\omega-8N_0a\left|\psi_0\right|^2\right)v-4N_0a\psi_0^{*2}u.
\end{align}
\label{eq:bdge}
\end{subequations}
A solution of these equations determines the behavior of a perturbation
$\delta\psi(x,t)=u(x)\exp(-\rmi \omega t)+v^*(x)\exp(\rmi \omega^* t)$ of a
normalized stationary state at interaction strength $N_0 a$. For real
frequencies $\omega$ the perturbed state performs stable oscillations around
the fixed point. If $\omega$ has a non-vanishing imaginary part it is necessary
to distinguish between two cases. A negative imaginary part describes an
exponentially damped and thus stable perturbation. By contrast, a perturbation
increases exponentially if the imaginary part is positive. If one or more
perturbations have a frequency with positive imaginary part the stationary
solution is unstable otherwise it is stable. Due to the ansatz of the
perturbation for every frequency $\omega$ with amplitudes $(u,v)$ a second
solution with frequency $-\omega^*$ and amplitudes $(v^*, u^*)$ exists.
Therefore all frequencies occur in pairs with positive and negative values of
$\real \omega$.  Applying the $\PT$ operator to the Bogoliubov-de Gennes
equations shows that if the stationary state $\psi_0$ has a perturbation
frequency $\omega$ then $\PT \psi_0$ has a perturbation frequency $\omega^*$.
Thus for $\PT$-symmetric eigenstates every frequency is always part of a set of
four solutions with $\pm \real \omega \pm \rmi \imag \omega$. For $\PT$-broken
eigenstates perturbations occur in pairs $\pm \real \omega + \rmi \imag \omega$
and since $\PT$-broken solutions are mapped onto each other by application of
the $\PT$ operator their perturbation frequencies are complex conjugate. 

\subsection{Stability of the $\PT$-symmetric solutions}
\begin{figure}%
  \centering%
  \includegraphics[width=\columnwidth]{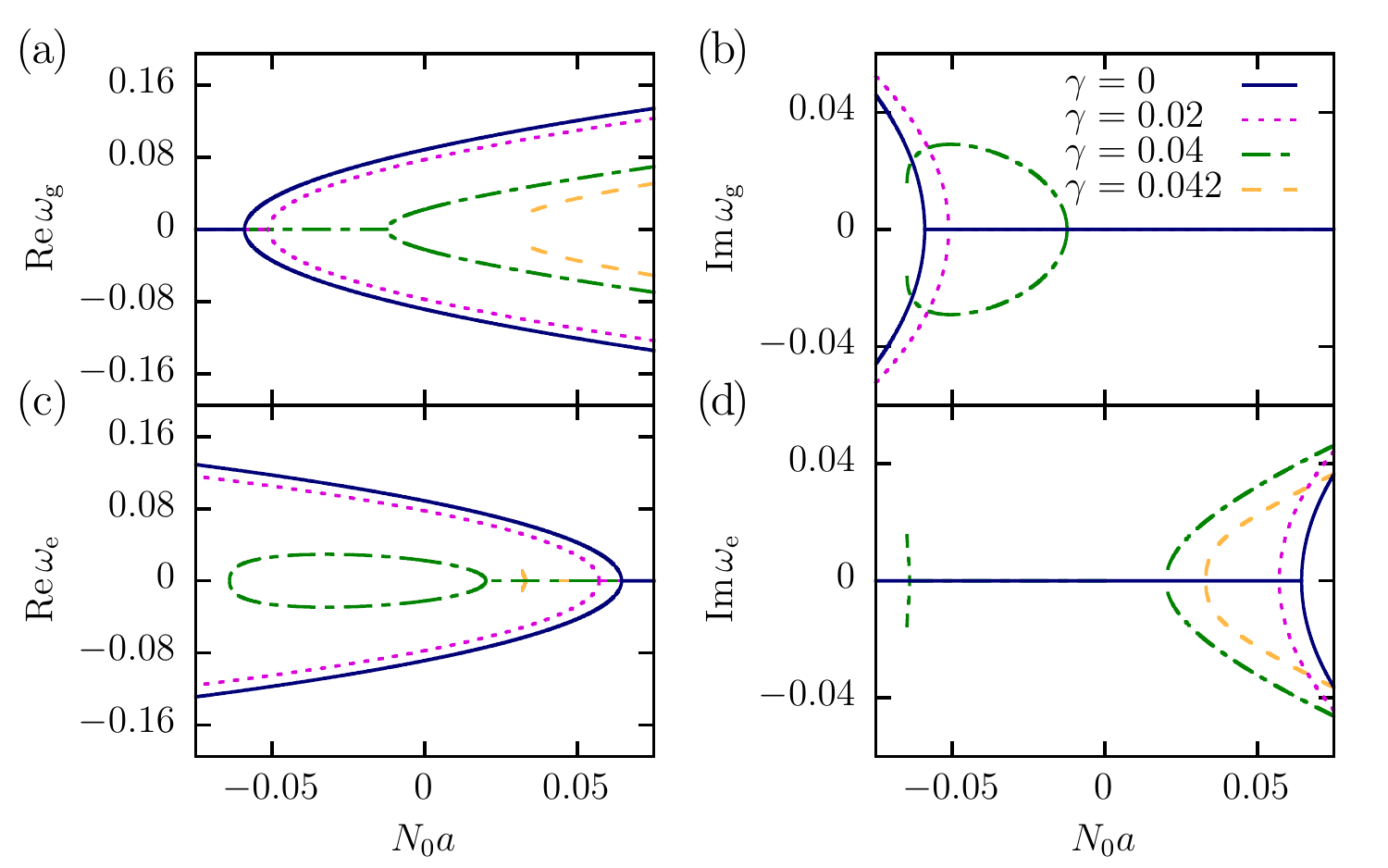}%
  \caption{%
    (Color online)
    (a),(c) Real and (b),(d) imaginary parts of the eigenvalues
    $\omega_\mrm{g/e}$ of the Bogoliubov-de Gennes equations for (a),(b) the
    ground state and (c),(d) the first excited state. Only the first nontrivial
    eigenvalue with smallest absolute real part is shown. Due to the symmetries
    of the equations the real and the imaginary parts of the eigenvalues both
    occur in pairs. Both states are stable for weak interactions but become
    unstable in the vicinity of the pitchfork bifurcations.
  }%
  \label{fig:stability_pt}%
\end{figure}

The Bogoliubov-de Gennes equations are solved numerically exact by integrating
the amplitudes $u$ and $v$ outwards and demanding that they vanish at the
boundaries. The first nontrivial Bogoliubov-de Gennes eigenvalue with smallest
absolute real part is shown in Fig.~\ref{fig:stability_pt} for the two
$\PT$-symmetric stationary states. Higher excitations are neglected since they
have real eigenvalues thus describing stable perturbations.

For a weak gain/loss contribution $\gamma=0.02$ the stability eigenvalues show
that both $\PT$-symmetric stationary states are stable for small interaction
strengths. The ground state (Fig.~\ref{fig:stability_pt}(a),(b)) becomes
unstable at attractive interactions and the first excited state
(Fig.~\ref{fig:stability_pt}(c),(d)) at repulsive interactions. Both stability
changes occur near the pitchfork bifurcations at which the $\PT$-broken states
emerge. For stronger gain/loss contributions $\gamma=0.04$ the pitchfork
bifurcations at which the two states become unstable are shifted to lower
absolute values of $N_0a$. Additionally we observe that the first excited
state becomes unstable for attractive interactions shortly before it merges
with the ground state at $N_0a\approx-0.065$ in a tangent bifurcation and
vanishes.

For gain/loss contributions $\gamma\gtrsim0.41$ the $\PT$-broken states exist
even for $N_0a=0$ and the $\PT$-symmetric stationary states exist only for
repulsive interactions. In this case the ground state is stable for all
interaction strengths whereas the excited state becomes unstable shortly after
it emerges in the tangent bifurcation at $N_0a\approx0.03$. It is worth noting
that the stability of the $\PT$-symmetric states changes not at the bifurcation
points but only in their vicinity. The reason for this behavior is discussed in
detail at the end of this section.

\subsection{Perturbations of the $\PT$-broken solutions}
\begin{figure}%
  \centering%
  \includegraphics[width=\columnwidth]{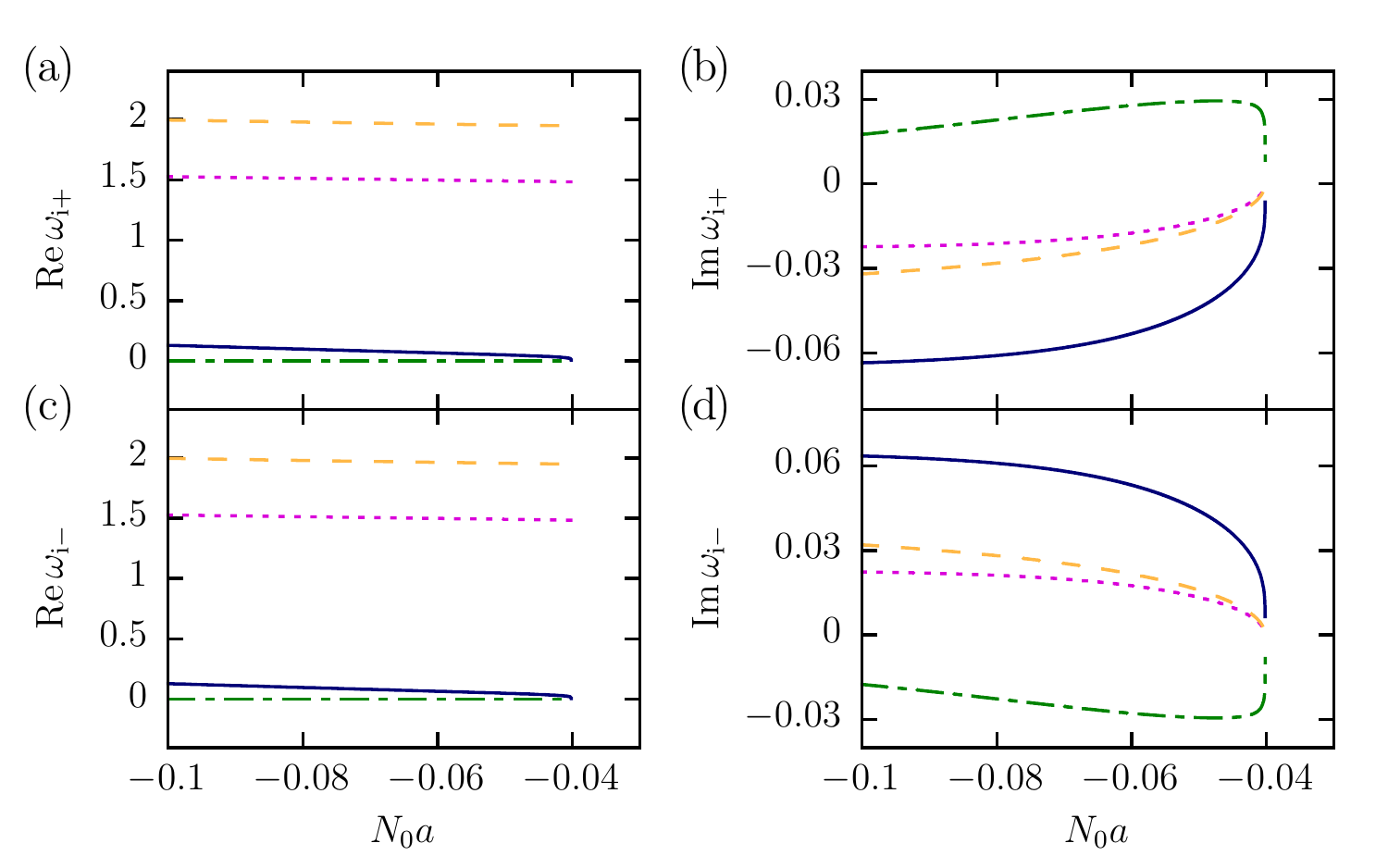}%
  \caption{%
    (Color online)
    (a),(c) Real and (b),(d) imaginary part of the Bogoliubov-de Gennes
    eigenvalues $\omega_{\rmi\pm}$ for the $\PT$-broken states with (a),(b)
    positive and (c),(d) negative imaginary part of the chemical potential
    $\mu$ at $\gamma=0.03$. The four eigenvalues with smallest real parts are
    shown. In addition to each eigenvalue $\omega$ a solution with
    negative real part $-\omega^*$ exists. All but one eigenvalue of the
    $\PT$-broken state with $\imag \mu > 0$ ($\imag \mu < 0$) have a negative
    (positive) imaginary part.
  }%
  \label{fig:stability_broken}%
\end{figure}

Although the $\PT$-broken solutions are no stationary solutions of the
time-dependent GPE it is nevertheless instructive to solve the Bogoliubov-de
Gennes equations for these states. We will see that this is relevant for the
understanding of the dynamics of the condensate as well as the stability of the
stationary $\PT$-symmetric solutions. Figure~\ref{fig:stability_broken} shows
the four Bogoliubov-de
Gennes eigenvalues with smallest absolute real part for the two $\PT$-broken
states using a constant value of the gain/loss parameter $\gamma=0.03$. The
stability eigenvalues of the $\PT$-broken state with $\imag \mu > 0$ in
Fig.~\ref{fig:stability_broken}(a),(b) show that the three eigenvalues with
nonzero real part have a negative imaginary part. In fact there is an infinite
number of further perturbations with negative imaginary part corresponding to
higher excited states of the double well. There is, however, an additional
solution with $\real \omega_{\rmi +} = 0$ and a positive imaginary part
describing a perturbation which increases exponentially. Since the
$\PT$-broken solutions are not stationary the usual interpretation of the
eigenvalues $\omega$ as stability indicators is invalid. The damped oscillatory
behavior described by the eigenvalues with negative imaginary part and
non-vanishing real part is characteristic for these states as will be seen in
Sec.~\ref{sec:dynamics}.

Since the $\PT$-broken states can be mapped onto each other by application of
the $\PT$ operator the $\PT$-broken state with $\imag \mu < 0$ has the complex
conjugate stability eigenvalues shown in
Fig.~\ref{fig:stability_broken}(c),(d). This is a consequence of the fact that
this state has the same dynamics as the $\PT$-broken state with $\imag \mu > 0$
if evolved in negative time direction.

\subsection{Stability at the bifurcation points}
\begin{figure}%
  \centering%
  \includegraphics[width=\columnwidth]{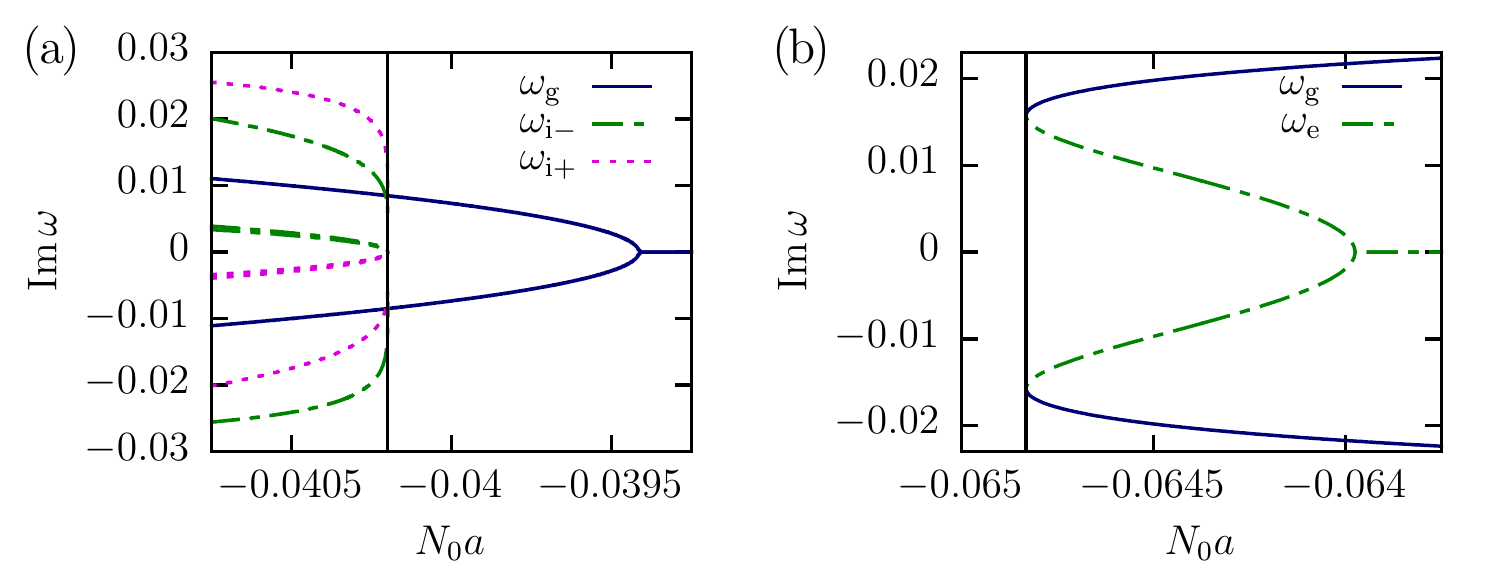}%
  \caption{%
    (Color online)
    Imaginary part of the Bogoliubov-de Gennes eigenvalues of all states
    involved in (a) the pitchfork bifurcation at $\gamma=0.03$ and (b) the
    tangent bifurcation at $\gamma=0.04$. The ground state becomes unstable
    before the pitchfork bifurcation at which the $\PT$-broken solutions
    arise. For decreasing $N_0a$ the excited state becomes unstable before the
    tangent bifurcation at which the two $\PT$-symmetric states coalesce and
    vanish. At the bifurcations the stability eigenvalues of the states
    involved coalesce since the states themselves become equal. The bifurcation
    points are marked by black vertical lines.
  }%
  \label{fig:gap}%
\end{figure}

Both the ground state and the excited state become unstable near a bifurcation
point. Therefore the Bogoliubov-de Gennes eigenvalues in the vicinity of the
bifurcations are now investigated in more detail. Figure~\ref{fig:gap} shows
the stability eigenvalues of all states involved in the bifurcations. It can
immediately be seen that the stability properties of the $\PT$-symmetric states
do not change at the bifurcation points but only in their vicinity. The ground
state (Fig.~\ref{fig:gap}(a)) is already unstable for greater values of $N_0a$,
i.e.\ the ground state becomes unstable in a parameter regime where the
$\PT$-broken solutions do not yet exist. This was already found in a two-mode
analysis of a $\PT$-symmetric double well \cite{Rodrigues13a}. A similar
behavior is observed at the tangent bifurcation where the excited state becomes
unstable shortly before the bifurcation point at which the $\PT$-symmetric
solutions vanish (Fig.~\ref{fig:gap}(b)). 

This discrepancy is surprising because we know from real nonlinear systems that
the stability of eigenstates changes at bifurcation points. Also in linear
$\PT$-symmetric systems all eigenstates are stable unless a $\PT$-broken
eigenstate with complex eigenvalue exists. In both cases a change in the
stability of a stationary state coincides with a qualitative change in the
spectrum. 

As a first step we ensure that the investigation of the lowest-lying
states is sufficient and higher excited states are not responsible for the
discrepancy. Therefore we solved the Bogoliubov-de Gennes equations for the
$\PT$-symmetric double-delta potential studied in~\cite{Cartarius12a,
Cartarius12b}, a system in which only two $\PT$-symmetric and two $\PT$-broken
eigenstates exist. Indeed, also this system shows the discrepancy thus ruling
out higher excited states as its origin \cite{Loehle14a}. It therefore seems
likely that the observed discrepancy is a consequence of the combination that
both a nonlinear and simultaneously $\PT$-symmetric system is investigated. Due
to the non-Hermiticity the norm is not conserved thus the nonlinearity
parameter $Na$ and consequently the spectrum changes with time. As a result the
dynamical properties are not governed by the eigenvalue spectrum at a fixed
interaction strength. Instead it is necessary to consider the whole spectrum as
a function of $Na$, and thus the $\PT$-broken states are already dynamically
accessible in the parameter regime where only $\PT$-symmetric states exist.

To confirm that this property is indeed the reason why the stability does not
change at the bifurcation points we modify the Gross-Pitaevskii nonlinearity
\begin{equation}
  |\psi(x,t)|^2 \to \frac{|\psi(x,t)|^2}{\int|\psi(x,t)|^2 \rmd x}.
  \label{eq:norm_independent_nonlin}
\end{equation}
This formulation is equivalent to the mean-field limit of the $\PT$-symmetric
Bose-Hubbard dimer by Graefe et al.\ in which such a discrepancy does not
occur~\cite{Graefe10a}.

Replacing the nonlinearity with~\eqref{eq:norm_independent_nonlin} does not
change the normalized eigenstates of the GPE. It leads, however, to a different
form of the Bogoliubov-de Gennes equations,
\begin{subequations}
\label{eq:bdge_ne}
\begin{align}
  \left(-\Delta+V-\omega-\mu-8N_0a\left|\psi_0\right|^2\right)u
    -4N_0av\psi_0^2 \notag\\
  +4N_0a\left|\psi_0\right|^2\psi_0\int v\psi_0+u\psi_0^*\rmd^3 r
    =0, \\
  \left(-\Delta+V^*+\omega-\mu^*-8N_0a\left|\psi_0\right|^2\right)v
    -4N_0au\psi_0^{*2} \notag\\
 +4N_0a\left|\psi_0\right|^2\psi_0^*\int v\psi_0+u\psi_0^*\rmd^3r=0.
\end{align}
\end{subequations}

\begin{figure}%
  \centering%
  \includegraphics[width=\columnwidth]{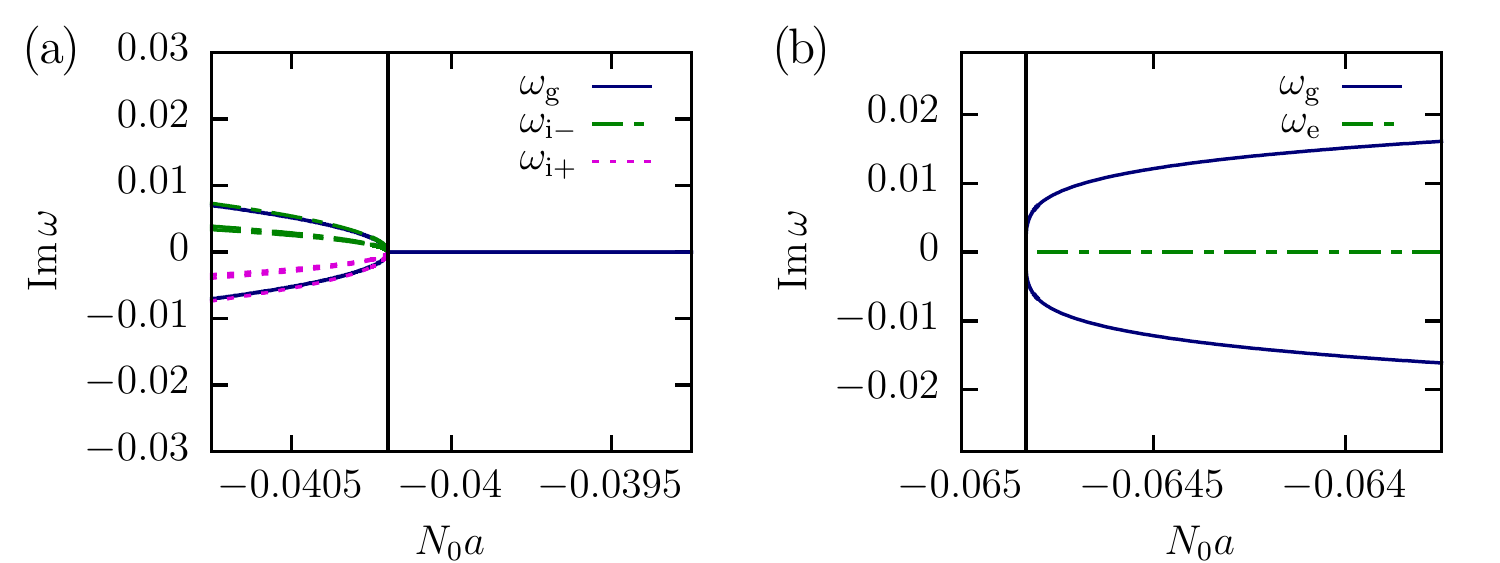}%
  \caption{%
    (Color online)
    Imaginary part of the stability eigenvalues $\omega$ of the Bogoliubov-de
    Gennes equations~\eqref{eq:bdge_ne} for the norm-independent
    nonlinearity~\eqref{eq:norm_independent_nonlin} in the vicinity of (a) the
    pitchfork bifurcation and (b) the tangent bifurcation. A constant
    gain/loss parameter (a) $\gamma=0.03$ and (b) $\gamma=0.04$ is used. The
    ground state becomes unstable at the pitchfork bifurcation and the excited
    state is stable until it vanishes at the tangent bifurcation. The
    $\PT$-broken states are now a pure sink or source. The bifurcation points
    are marked by black vertical lines.
  }%
  \label{fig:gap_hold}%
\end{figure}

These equations are solved in the vicinity of the bifurcations so that the
results with the adapted nonlinearity~\eqref{eq:norm_independent_nonlin} in
Fig.~\ref{fig:gap_hold} can directly be compared to the stability eigenvalues
obtained with the usual Gross-Pitaevskii nonlinearity in Fig.~\ref{fig:gap}.
Using the adapted nonlinearity~\eqref{eq:norm_independent_nonlin} the stability
of the ground state changes at the bifurcation point, i.e.\ the ground state
becomes unstable as soon as the $\PT$-broken solutions exist. Additionally the
excited state does not show a stability change for attractive interaction but
stays stable until it vanishes. Thus, the behavior observed that the stability
of the $\PT$-symmetric states does not change at the bifurcation points but
only in their vicinity is indeed a result of the norm-dependent nonlinearity of
the GPE. 

Also the behavior of the $\PT$-broken states does change by introducing the
adapted nonlinearity. With the nonlinearity~\eqref{eq:norm_independent_nonlin}
the $\PT$-broken states are eigenstates of the time-dependent GPE with
exponentially increasing or decreasing norm proportional to $\exp(2\imag\mu)$.
Furthermore the $\PT$-broken states with positive and negative imaginary part
are now a pure sink or source, respectively.

\section{Wave packet dynamics} 
\label{sec:dynamics}
The linear stability analysis in the previous section describes the dynamical
behavior in the vicinity of the eigenstates, and in the case of the
$\PT$-broken solutions for very short time spans after which the spectrum
changes.  For finite time spans this is not sufficient. To gain a more
elaborate picture of the dynamics the time evolution of wave packets is
investigated for different values of the gain/loss parameter $\gamma$ and the
nonlinearity parameter $Na$.

We use the \textit{split-operator method} to numerically calculate the time
evolution of wave packets which is known to produce accurate results even for
nonlinear equations as the GPE~~\cite{Feit82a, Fleck76a, Javanainen06a}.

The oscillation of a single wave packet between the wells of the
$\PT$-symmetric double-well potential has already been discussed
in~\cite{Dast12a}. There the square modulus of the wave packets was
investigated to analyze the characteristic phase shift of the oscillations.
However, this representation does not allow for predictions of the behavior of
arbitrary wave packets and the impact of the eigenstates of the system. We will
now choose the Bloch sphere formalism as a different approach to visualize the
time evolution of arbitrary states.

Even though Bloch sphere representations have already been used for
$\PT$-symmetric~\cite{Graefe08b, Graefe10a} as well as
dissipative~\cite{Trimborn08a} two-mode BECs these studies always restricted
the dynamics to the surface of the Bloch sphere. In non-Hermitian nonlinear
systems, in which the norm is not conserved and the associated differential
equation depends explicitly on the norm of the wave function, the dynamics
is in general not restricted to this surface. We will see that the Bloch
sphere provides a significant insight into the dynamical properties
nonetheless.

\subsection{Bloch sphere formalism}
In general the representation as a Bloch sphere is limited to two-level quantum
systems. Since the Hilbert space of the system investigated is not
two-dimensional we use a projection to the space spanned by the $\PT$-symmetric
ground state $\psi_\mrm{g}$ and excited state $\psi_\mrm{e}$. In the linear
case $Na=0$ the time evolution of initial wave functions consisting of a linear
superposition of the ground and excited state is restricted to this
two-dimensional space. With interaction $Na\neq0$ this is no longer true,
however, our calculations show that for all time evolutions considered the
projection to two dimensions is still a very good approximation.

We now choose two orthogonal basis vectors of the space spanned by the two
$\PT$-symmetric stationary solutions. The first basis vector is identical to
the normalized ground state $\ket{e_1} = \ket{\psi_\mrm{g}}$, and the second
basis vector, $\ket{e_2} = \alpha(\ket{\psi_\mrm{e}} -
\braket{\psi_\mrm{g}}{\psi_\mrm{e}} \ket{\psi_\mrm{g}})$ with normalization
constant $\alpha$, is the component of the excited state $\ket{\psi_\mrm{e}}$
orthogonal to $\ket{e_1}$ and is selected by application of the Gram-Schmidt
method. Both basis vectors are exactly $\PT$-symmetric. An arbitrary wave
function can be written as
\begin{equation} 
  \ket{\psi} = c_1 \ket{e_1} + c_2 \ket{e_2} + \ket{\psi_\mrm{err}}
  \label{eq:bloch_projection}
\end{equation}
with $c_1 = \braket{e_1}{\psi}$ and $c_2 = \braket{e_2}{\psi}$. Although all
initial wave packets considered are superpositions of $\ket{e_1}$ and
$\ket{e_2}$ the time evolution will in general leave the space spanned by
$\ket{e_1}$ and $\ket{e_2}$. The norm of $\ket{\psi_\mrm{err}}$ measures the
error made by the projection to the two-dimensional Hilbert space. In all
calculations presented $\braket{\psi_\mrm{err}}{\psi_\mrm{err}} < 0.004$ was
found, justifying the projection.

\begin{figure}%
  \centering%
  \includegraphics[width=0.67\columnwidth]{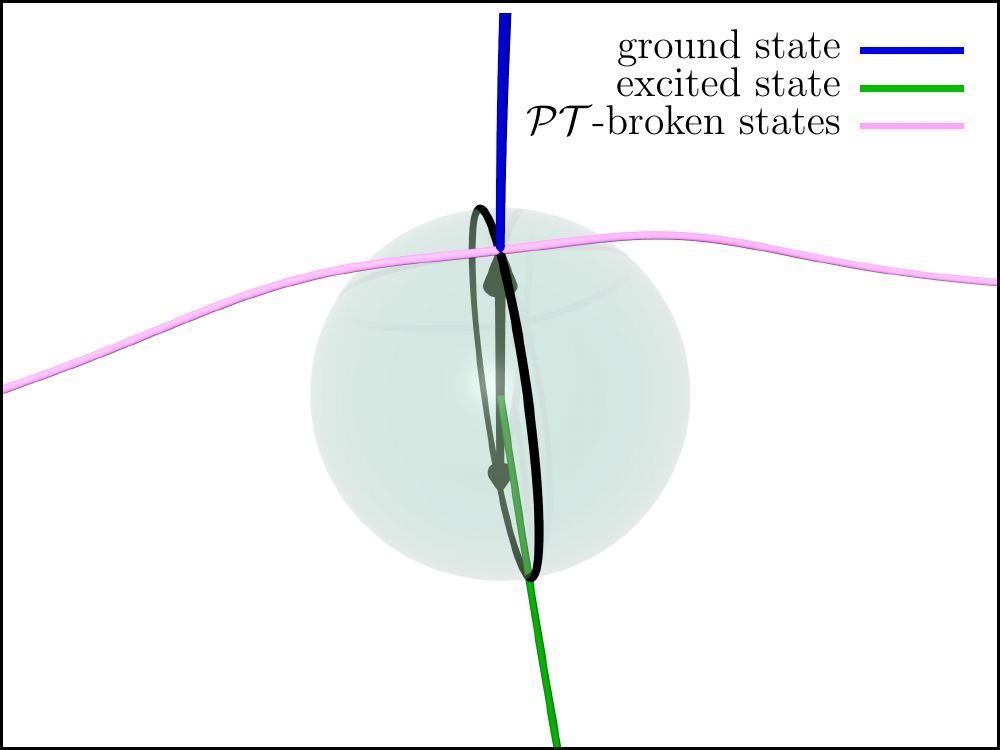}%
  \caption{%
    (Color online)
    The Bloch sphere representation using the coordinates defined in
    Eq.~\eqref{eq:bloch_coordinates}. The basis vectors $\ket{e_1}$ and
    $\ket{e_2}$ correspond to the north and south pole, respectively. On the
    front side of the plotted great cycle (black line) the azimuth angle is
    $\varphi=0$ and on the back side $\varphi=\pi$. All states residing on the
    plane in which this great cycle lies are $\PT$-symmetric and the time
    evolution in the system is symmetric with respect to this plane.
    Furthermore the great cycle is used as the starting point for trajectories
    discussed in Sec.~\ref{sec:bloch_dynamics}. Additionally the four
    eigenstates are plotted for $\gamma=0.03$ and the value of $Na$
    corresponding to the current location 
    (cf.\ Sec.~\ref{sec:bloch_eigenstates}).
  }%
  \label{fig:bloch_orientation}%
\end{figure}

The basis vectors $\ket{e_1}$ and $\ket{e_2}$ are defined to correspond to the
north and south pole of the Bloch sphere, respectively, by introducing the
spherical coordinates $R\in[0,\infty)$, $\varphi\in(-\pi,\pi]$ and
$\vartheta\in[0,\pi]$ as follows
\begin{subequations}
  \label{eq:bloch_coordinates}
  \begin{align}
    c_1 &= R \, \rme^{\rmi\chi+\rmi{\varphi}} \cos(\vartheta/2), \\
    c_2 &= R \, \rme^{\rmi\chi} \sin(\vartheta/2),
  \end{align}
\end{subequations}
with an arbitrary phase $\chi\in\mathbb{R}$. Since
$\braket{\psi_\mrm{err}}{\psi_\mrm{err}} \ll 1$ the radius can be identified
with the norm of the wave packet $R \approx ||\psi||$. The orientation of the
basis vectors on the Bloch sphere and the stationary solutions at a fixed value
of $\gamma$ are shown in Fig.~\ref{fig:bloch_orientation}.

The $\PT$ symmetry of the system has several implications for the
representation as a Bloch sphere. Since the system considered is non-Hermitian
the norm of a wave packet is not conserved. The norm was identified with the
radius $R$, thus in general the time evolution of an arbitrary wave packet is
not constrained to the surface of the Bloch sphere but it will either dive into
the sphere or leave the surface to larger radii. The two basis vectors are
exactly $\PT$-symmetric thus application of the $\PT$ operator leads to a
complex conjugation of the coefficients $c_1$ and $c_2$ which is equivalent to
the reflection $\varphi \to -\varphi$. Consequently all $\PT$-symmetric states
reside on the plane defined by $\varphi\in\{0,\pi\}$.

A further implication of the $\PT$ symmetry of the system is that if
$\psi(x,t)$ is a solution of the time-dependent GPE then $\psi^*(-x,-t)$ is
also a solution. Since $\mathcal{P}$ reflects the spatial coordinate and
$\mathcal{T}$ applies only a complex conjugation $\psi^*(-x,-t)=\PT\psi(x,-t)$
holds and hence all trajectories are symmetric with respect to the plane
$\varphi\in\{0,\pi\}$, in which the $\PT$ symmetric eigenstates are found. 

\subsection{Eigenstates in Bloch sphere representation}
\label{sec:bloch_eigenstates}
\begin{figure*}%
  \centering%
  \includegraphics[width=\textwidth]{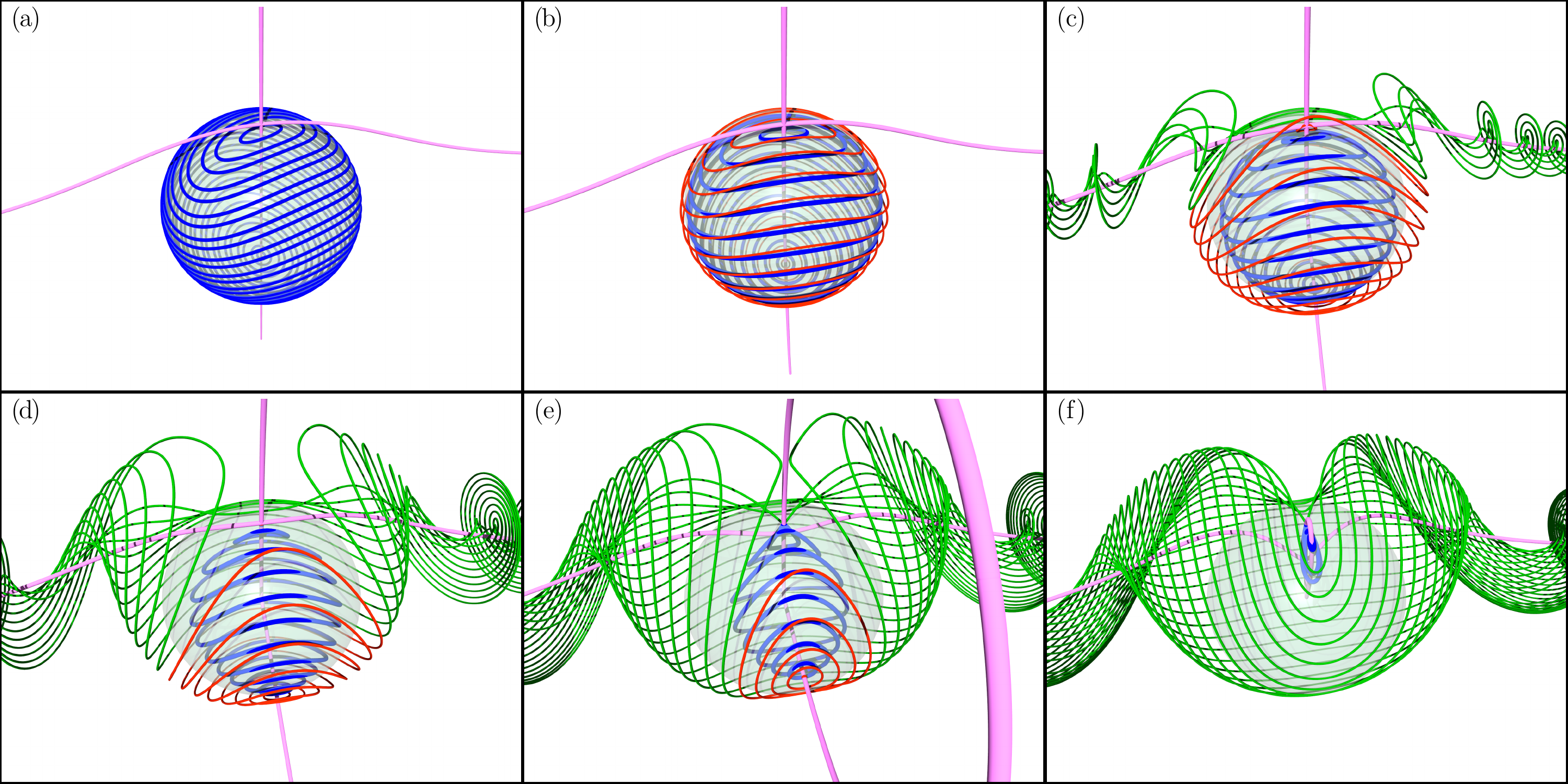}%
  \caption{%
    (Color online)
    The dynamics of wave packets projected to the space spanned by the
    $\PT$-symmetric ground and excited state illustrated on a Bloch sphere. The
    interaction strength $Na=-0.05$ on the surface of the sphere is constant in
    all figures whereas the gain/loss parameters are varied, with (a)
    $\gamma=0$, (b) $\gamma=0.0025$, (c) $\gamma=0.01$, (d) $\gamma=0.02$, (e)
    $\gamma=0.03$, and (f) $\gamma=0.04$. The solutions of the time-independent
    GPE are plotted for orientation (pink lines
    cf.~Fig.~\ref{fig:bloch_orientation}). The ground state starts in the
    center of the sphere and goes through the north pole whereas the
    penetration point of the excited state is at the south pole for $\gamma=0$
    and wanders on the meridian $\varphi=0$ to the north pole for increasing
    values of $\gamma$.  The $\PT$-broken solutions emerge from the ground
    state. All wave packets shown start on a great cycle through the north
    pole, south pole and the excited state. Wave packets starting on the
    meridian between the ground state at the north pole and the excited state
    on the front side of the sphere evolve to a smaller norm thus diving into
    the sphere (closed thick blue lines inside the sphere). Wave packets
    starting in the remaining region of the great circle either show
    oscillations with a larger norm outside the sphere (closed red lines
    outside the sphere) or diverge while encircling the $\PT$-broken solutions
    (green lines departing from the left and the right side).
  }%
  \label{fig:bloch_nonlin}%
\end{figure*}

Figure~\ref{fig:bloch_nonlin} shows both the eigenstates and time-evolved wave
packets in the Bloch sphere representation. In this example the Bloch sphere
represents the interaction strength $Na=N_0a=-0.05$. Since a larger radius is
equivalent to a greater amount of particles, $N= N_0||\psi||^2 = N_0 R^2$,
plotting the $\PT$-symmetric and $\PT$-broken solutions of the time-independent
GPE is another way of showing these eigenstates in dependence of $Na$. They are
depicted as thick pink lines in Fig.~\ref{fig:bloch_nonlin} and can be seen
most clearly in Fig.~\ref{fig:bloch_nonlin}(a). Only attractive interactions
are shown thus larger radii relate to more negative values of $Na$.

Both $\PT$-symmetric solutions start at the center of the sphere. The ground
state goes through the north pole and the excited state goes through the south
pole for $\gamma=0$ (Fig.~\ref{fig:bloch_nonlin}(a)). Increasing $\gamma$ the
penetration point of the excited state through the Bloch sphere wanders on the
meridian $\varphi=0$ to the north pole at which the ground state resides
(cf.~Figs~\ref{fig:bloch_nonlin}(b)-(e)). For a critical value of $\gamma$ this
point reaches the north pole, which is almost fulfilled in
Fig.~\ref{fig:bloch_nonlin}(f). For greater values of $\gamma$ the
$\PT$-symmetric solutions even vanish on the surface of the Bloch sphere and
only exist within the sphere. This behavior of the stationary solutions on the
surface of the Bloch sphere can be comprehended by comparison with the
eigenvalue spectrum in Fig.~\ref{fig:spectrum} for different values of $\gamma$
and a fixed value of $Na$.

As already shown in the eigenvalue spectrum the bifurcation at which the
ground and excited state coalesce can also be reached by tuning $Na$. This can
be seen in Fig.~\ref{fig:bloch_nonlin}(e) where the two $\PT$-symmetric states
coalesce at a critical radius outside the Bloch sphere resulting in the
closed circle of the thick pink lines. For smaller values of $\gamma$ shown in
Fig.~\ref{fig:bloch_nonlin}(a)-(d) the bifurcation point lies at a larger
radius outside the figure.

Since attractive interactions are shown the two $\PT$-broken solutions emerge
from the ground state. For the chosen interaction strength on the Bloch sphere
$Na=-0.05$ the $\PT$-broken solutions emerge outside the sphere for
$\gamma\gtrsim0.022$ and inside the sphere for $\gamma\lesssim0.022$. The
$\PT$-broken solution with $\imag\mu>0$ ($\imag\mu<0$) lies on the left (right)
side of the symmetry plane.

\subsection{Dynamics on the Bloch sphere}
\label{sec:bloch_dynamics}
After this short discussion of the eigenstates we now address the time
evolution. For all calculations shown in Fig.~\ref{fig:bloch_nonlin} the
initial wave packets are normalized and a linear superposition of $\ket{e_1}$
and $\ket{e_2}$, thus $R=1$ and $\braket{\psi_\mrm{err}}{\psi_\mrm{err}}=0$.
The azimuth angle is either $\varphi=0$ or $\pi$, therefore all initial wave
packets start on a great circle of the Bloch sphere through the north pole,
south pole, and the excited stationary state. They are integrated in positive
and negative time direction.

For the case $\gamma=0$ shown in Fig.~\ref{fig:bloch_nonlin}(a)
the norm and therefore the radius are conserved quantities. All trajectories
stay on the surface of the Bloch sphere. Both the stationary ground and excited
state are elliptic fixed points and therefore stable.

This behavior changes drastically if a small gain/loss parameter
$\gamma=0.0025$ is introduced as done in Fig.~\ref{fig:bloch_nonlin}(b). Due to
the gain and loss, the norm of the wave packets are no longer conserved and the
trajectories no longer run on the surface of the Bloch sphere. We identify two
different regions on the great circle for initial wave packets which are
delimited by the ground state, viz.\ the north pole, and the excited state. All
wave packet starting in the region on the front side between the two fixed
points (thick blue lines) evolve to a smaller norm inside the Bloch sphere
whereas wave packets starting on the second region on the back side (red lines)
evolve to a higher norm outside the sphere. The two $\PT$-symmetric stationary
states are again elliptic fixed points, thus being stable. The sum of all
oscillating trajectories of the wave packets define closed surfaces which
cannot be penetrated by other trajectories.

Increasing the gain/loss parameter to $\gamma=0.01$ leads to the situation
shown in Fig.~\ref{fig:bloch_nonlin}(c). Again the wave packets starting on the
front region between the two fixed points oscillate to a smaller norm and
define a closed surface inside the Bloch sphere. However, the wave packets
starting on the back region and oscillating outside the Bloch sphere do no
longer define a closed surface. Instead an additional type of trajectories with
diverging norm encircling the $\PT$-broken eigenstates appears. For
$\gamma=0.02$ the amount of diverging trajectories increases as can be seen in
Fig.~\ref{fig:bloch_nonlin}(d). These trajectories arrive from the vicinity of
the $\PT$-broken state with $\imag\mu<0$ on the right side, touch the sphere,
and leave the region of the sphere to larger radii $R$ encircling the path of
the $\PT$-broken state with $\imag\mu>0$ on the left side. This illustrates the
role of the $\PT$-broken solutions for the dynamics of the condensate as sink
and source.

Another qualitative change in the dynamical behavior is found for $\gamma=0.03$
in Fig.~\ref{fig:bloch_nonlin}(e). In agreement with the investigation of the
linear stability in Fig.~\ref{fig:stability_pt} the ground state is unstable
for $Na=-0.05$, viz.\ exactly on the surface of the Bloch sphere. For lower
values of $\gamma$ there is a region around the ground state in which only
stable oscillations originate. For $\gamma=0.03$ the ground state is unstable
and all wave packets starting on the $\varphi=\pi$ meridian behind the ground
state are diverging. Wave packets starting before the ground state still show
stable oscillations evolving inside the sphere.

Finally for $\gamma=0.04$ shown in Fig.~\ref{fig:bloch_nonlin}(f) the two
$\PT$-symmetric stationary solutions are almost identical and most wave packets
starting on the great circle on the Bloch sphere diverge. Only wave packets
starting in a small region around the stable excited state and in the region
between the two stationary states still show stable oscillations. For even
greater values of $\gamma$ the stationary solutions do no longer exist on the
surface of the Bloch sphere and no stable oscillations starting on the great
circle can be observed.

\section{Conclusion and outlook}
\label{sec:conclusion}
We studied the implications of $\PT$ symmetry on the dynamical behavior and
stability of a BEC with contact interaction in a
double-well potential.

Solving the Bogoliubov-de Gennes equations for non-Hermitian systems showed
that the two $\PT$-symmetric stationary solutions with smallest chemical
potential are stable as long as the $\PT$-broken states do not exist.
However, the ground state becomes dynamically unstable in the vicinity of the
pitchfork bifurcation at which the $\PT$-broken states emerge from the ground
state at an attractive interaction strength. Analogously the excited state
becomes unstable at repulsive interactions in the vicinity of the pitchfork
bifurcation. The discrepancy between the bifurcations and the points at which
the stability changes could be traced back to the norm dependency of the
Gross-Pitaevskii nonlinearity.

Due to the non-Hermiticity of the system the dynamics is not governed by
isolated fixed points but an infinite number of eigenstates which solve
the time-independent GPE for the varying number of particles. The dynamics of
the condensate was visualized using the Bloch sphere formalism although the
dynamics is not constrained to the surface of the sphere due to the
nonlinearity and non-Hermiticity of the system. Applying a small gain and loss
of particles leaves the typical dynamics in a real double-well potential mostly
intact but the trajectories now run slightly above or below the surface of the
sphere describing a condensate with more or less particles,
respectively. For stronger gain/loss contributions an additional type of
trajectories arises which describes a condensate localized in one well with
a diverging number of particles. These diverging trajectories encircle the
$\PT$-broken eigenstates of the time-independent GPE. However it is still
possible to choose initial wave packets that show stable oscillations. In fact
we observed that all $\PT$-symmetric wave functions which initially dive into
the sphere always show stable oscillations. If the gain and loss is further
increased most wave packets diverge and stable oscillations are only found in a
small region in the vicinity of the excited state.

Understanding the dynamics of a BEC in a $\PT$-symmetric double well is the
first step towards an experimental realization of a $\PT$-symmetric quantum
system and the starting point for studies in more complex potentials and with
additional interaction types like the dipolar interaction. Additionally it
would be highly desirable to obtain a microscopic description of the in- and
outcoupling process represented by an imaginary potential in the mean-field
limit. Analyzing the bifurcation scenario at strong attractive interaction
strengths switching from two tangent bifurcations to one in the presence of
gain and loss as visible in Fig.~\ref{fig:spectrum_3d} is an interesting task
for future work from a more theoretical point of view and can probably be
achieved using the analytic continuation described in~\cite{Dast13a}.

\end{document}